\documentclass[twocolumn,showpacs,preprintnumbers]{revtex4}%
\usepackage{amssymb}
\usepackage{amsfonts}
\usepackage{amsmath}
\usepackage{graphicx}
\usepackage{dcolumn}
\usepackage{bm}%
\begin{document}
\preprint{ }
\title{Signal Amplification in NbN Superconducting Resonators via Stochastic
Resonance }
\author{Baleegh Abdo}
\email{baleegh@tx.technion.ac.il}
\author{Eran Arbel-Segev}
\author{Oleg Shtempluck}
\author{Eyal Buks}
\affiliation{Department of Electrical Engineering, Technion, Haifa 32000, Israel}
\date{\today}

\begin{abstract}
We exploit nonlinearity in NbN superconducting stripline resonators, which is
originated by local thermal instability, for studying stochastic resonance. As
the resonators are driven into instability, small amplitude modulated (AM)
microwave signals are amplified with the aid of injected white noise. The
dependence of the signal amplification on the modulation amplitude and the
modulation frequency is examined and compared with theory.

\end{abstract}
\maketitle




The notion that certain amount of white noise can appreciably amplify small
periodic modulating signals acting on bistable systems, generally known as
stochastic resonance, has been over the last two decades of a great interest
\cite{SR review,Tuning in to noise,Bona Fide,Amplification,Amp dist}. It has
been applied for instance to account for the periodicity of ice ages occurring
on earth by Benzi and collaborators \cite{Benzi SR,Benzi Ice}. It was
demonstrated experimentally in several electronic devices such as ac-driven
Schmitt triggers \cite{schmitt}. It has a growing role in explaining some
important neurophysiological processes in neuronal systems
\cite{neuronal,cricket}. In addition, it was used to amplify small signals in
various nonlinear systems, e. g. the intensity of one laser mode in a bistable
ring laser \cite{ring}, the magnetic flux in a superconducting quantum
interference device \cite{flux,JJloop}, and even more recently, a small
periodic drive of a nanomechanical oscillator \cite{Beams}. In this work we
exploit the metastability exhibited by our nonlinear NbN superconducting
stripline resonators in order to amplify a small applied sinusoidal modulating
signal using stochastic resonance. Moreover, we apply stochastic resonance
measurements in order to examine the dependence of the signal amplification
generated by the resonator system on both the modulation frequency and amplitude.

\begin{figure}
[ptb]
\begin{center}
\includegraphics[
height=2.4561in,
width=3.3615in
]%
{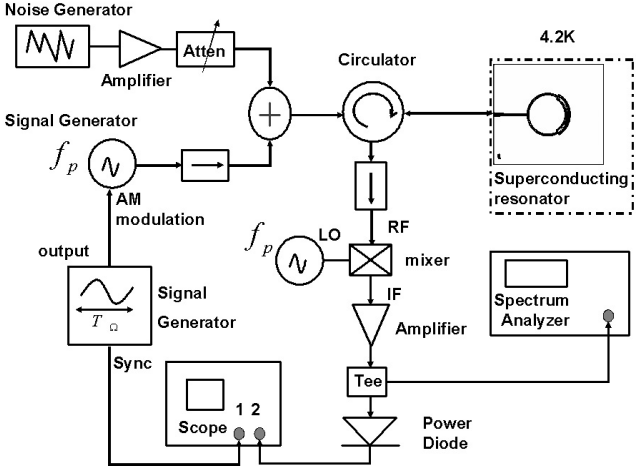}%
\caption{Schematic drawing of the experimental setup used to measure
stochastic resonance. The microwave signal generator and the local oscillator
at frequency $f_{p}$ were phase locked. The layout of the resonator is shown
at the top-right corner.}%
\label{setup}%
\end{center}
\end{figure}

The layout of the center layer of the NbN stripline resonator that was used is
shown at the top-right corner of Fig. \ref{setup}. Fabrication details as well
as nonlinear characterization and modeling of these resonators can be found in
Refs \cite{observation,dynamics}.%

\begin{figure}
[ptb]
\begin{center}
\includegraphics[
height=3.1773in,
width=3.5518in
]%
{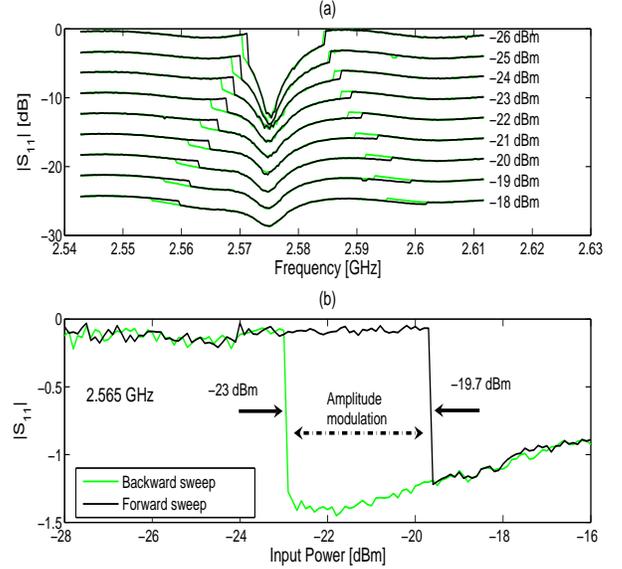}%
\caption{(Color online). (a) Forward and backward frequency sweeps applied to
the first mode of the resonator at $\sim2.57\operatorname{GHz}.$ The sweeps
exhibit hysteresis loops at both sides of the resonance line shape. The plots
corresponding to different input powers were shifted by a vertical offset for
clarity. (b) Reflected power hysteresis measured at a constant angular
frequency of $\omega_{p}=2\pi\cdot2.565\operatorname{GHz}$ which resides
within the left-side metastable region of the resonance. For both plots the
black (dark) line represents a forward sweep whereas the green (light) line
represents a backward sweep.}%
\label{frequencyhysteresisAndPower}%
\end{center}
\end{figure}

As it was mentioned earlier, one of the necessary conditions for the
demonstration of stochastic resonance in a nonlinear system is the existence
of metastable states, which the system can hop between with the aid of
stochastic noise and a suitable\textit{ }modulation drive. Thus, in order to
set a possible working point of the resonator at the metastable region, two
preliminary hysteresis measurements were performed. In one measurement
exhibited in Fig. \ref{frequencyhysteresisAndPower} (a), forward and backward
frequency sweeps of the reflection parameter $S_{11}$ were measured for the
fundamental mode of the resonator at $f_{0}\simeq2.57%
\operatorname{GHz}%
$. The bidirectional frequency sweeps form two hysteresis loops at both sides
of the resonance line shape at which the resonator becomes bistable. In
another measurement shown in Fig. \ref{frequencyhysteresisAndPower} (b) the
frequency of the pump $f_{p}$ was set to $2.565%
\operatorname{GHz}%
$ positioned at the left side of the resonance, while the input power was
swept in the forward and the backward directions. A hysteresis loop of the
reflection parameter is apparent in this measurement as well, this time along
the power axis. Thus, the working point was set to $f_{p}=2.565%
\operatorname{GHz}%
,$ $P_{0}=-21.5$ dBm, while the applied modulation drive was a sinusoidal AM
signal with a modulation amplitude $A_{\operatorname{mod}}=0.27$.%

A schematic diagram of the experimental setup employed to measure
\textit{stochastic resonance} is depicted in Fig. \ref{setup}. A coherent
signal $P_{0}\cos(\omega_{p}t)$ with angular frequency $\omega_{p}=2\pi\cdot
f_{p}$ is AM modulated using a sinusoidal generator with an angular frequency
$\Omega.$ The modulated signal is combined with white noise and injected into
the resonator. The white noise which is generated using a noise source is
amplified using an amplifying stage and tuned via an adjustable attenuator.
The reflected power off the resonator is mixed with a local oscillator with
frequency $f_{p}$ and measured simultaneously in the time and frequency
domains via an oscilloscope and a spectrum analyzer respectively. Thus, the
input signal power fed to the test port of the resonator (after calibrating
the path losses) reads%

\begin{equation}
P_{in}(t)=P_{0}\left[  1+A_{\operatorname{mod}}\sin\left(  \Omega
t+\varphi\right)  \right]  \cos(\omega_{p}t)+\xi(t), \label{Pint}%
\end{equation}
where $\xi(t)$ denotes a zero-mean Gaussian white noise $\left\langle
\xi(t)\right\rangle =0,$ with autocorrelation function $\left\langle \xi
(t)\xi(t^{\prime})\right\rangle =2D\delta(t-t^{\prime}),$ where $D$ is the
noise intensity. For convenience, we choose the phase of the periodic drive to
be zero $\varphi=0.$

For small amplitudes of the modulation signal $A_{\operatorname{mod}}\ll1$ and
in steady state conditions, the reflected power off the resonator $P_{r}(t)$
measured via a homodyne detection setup at the output, can in general, be
written in the form \cite{Amplification,Amp dist}%

\begin{equation}
P_{r}(t)=P_{0}^{r}+\sum\limits_{n=1}^{\infty}A_{n}^{r}(D)\cos\left(  n\Omega
t+\phi_{n}(D)\right)  +\xi^{r}(t), \label{Prt}%
\end{equation}
where $P_{r}(t)$ is a superposition of a dc component $P_{0}^{r},$ periodic
functions of time with integer multiples of the fundamental angular frequency
$\Omega$, and $\xi^{r}(t)$ the noise at the output. The coefficients
$A_{n}^{r}(D)$ and $\phi_{n}(D)$ represent the reflected power amplitude and
the phase lag of the \textit{n}-th harmonic as a function of the noise
intensity respectively. Calculating the phase-averaged spectral density of
$P_{r}(t)$ yields%

\begin{align}
S(\omega)  &  =2\pi P_{0}^{r}\delta(\omega)\label{Sw}\\
&  +\pi\sum\limits_{n=1}^{\infty}A_{n}^{r}(D)[\delta\left(  \omega
-n\Omega\right)  +\delta\left(  \omega+n\Omega\right)  ]+S_{N}(\omega
),\nonumber
\end{align}
which is composed of a delta spike at dc ($\omega=0)$, delta spikes with
amplitudes $A_{n}^{r}(D)$ centered at $\omega=\pm n\Omega$, $n=1,2,3...,$ and
a background spectral density of the noise denoted by $S_{N}(\omega).$

As it is known, one of the distinguished fingerprints of stochastic resonance
phenomenon is a peak observed in the signal to noise ratio (SNR) curve as a
function of the input noise intensity $D,$ corresponding to some nonzero
intensity $D_{\mathrm{SR}}.$ This counterintuitive amplification in SNR curve
is generally explained in terms of coherent interaction between the modulating
signal and the stochastic noise entering the system.

In this framework a SNR for the \textit{n}-th harmonic can be defined as
\cite{SR review}%

\begin{equation}
\mathrm{SNR}_{n}=2\left(  \lim_{\Delta\omega\rightarrow0}\int\limits_{n\Omega
-\Delta\omega}^{n\Omega+\Delta\omega}S\left(  \omega\right)  d\omega\right)
/S_{N}\left(  n\Omega\right)  =\frac{2\pi A_{n}^{r}(D)}{S_{N}\left(
n\Omega\right)  }, \label{snr}%
\end{equation}
where SNR of the fundamental harmonic corresponds to $n=1.$%

\begin{figure}
[ptb]
\begin{center}
\includegraphics[
height=2.8418in,
width=3.5103in
]%
{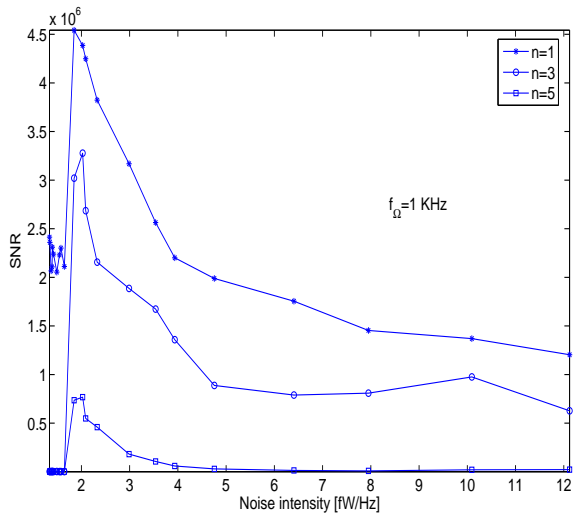}%
\caption{SNR curves measured for the fundamental $n=1$ and the odd harmonics
$n=3$, $n=5,$ as a function of the input noise intensity. A peak in the
$\ \mathrm{SNR}$ is detected at around $D_{\mathrm{SR}}$ of $2$
f$\operatorname{W}/\operatorname{Hz}$. }%
\label{SNR}%
\end{center}
\end{figure}

In Fig. \ref{SNR}, three SNR curves as a function of noise intensity are
shown, corresponding to the fundamental $n=1$ and the odd harmonics $n=3$ and
$n=5.$ All three curves display a synchronized peak in the SNR around
$D_{\mathrm{SR}}$ of about $2$ fW/Hz.%

\begin{figure}
[ptb]
\begin{center}
\includegraphics[
height=3.3347in,
width=3.6357in
]%
{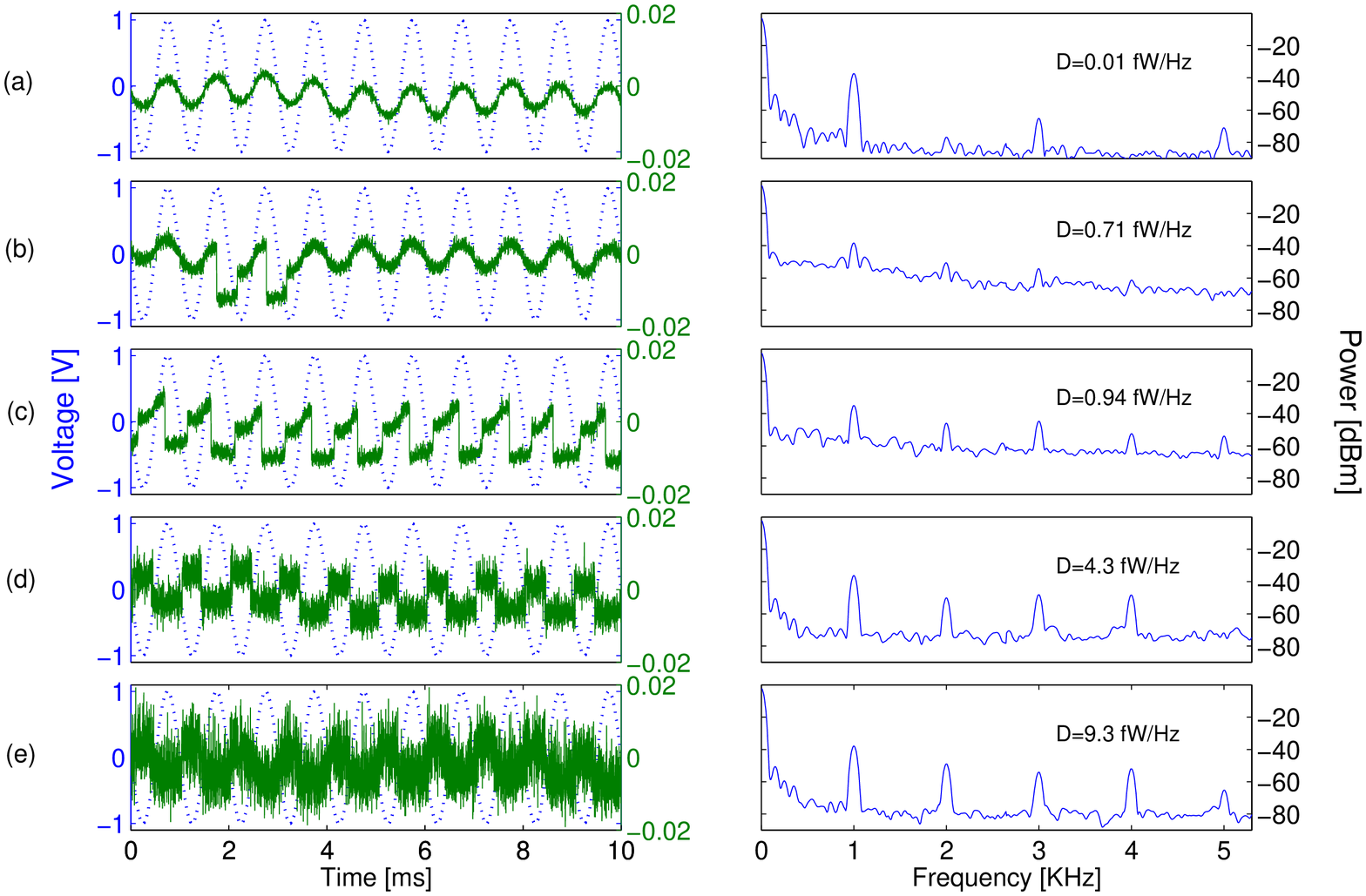}%
\caption{(Color online). Panels (a)-(e) exhibit typical snapshots of the
reflected modulated signal in the time domain (left side) and in the frequency
domain (right side) as the input noise intensity $D$ is increased. Panels
(a)-(b) correspond to noise intensities below $D_{\mathrm{SR}}.$ Panel (c)
corresponds to a noise intensity of $D_{\mathrm{SR}}=0.94$ fW/Hz. Panels
(d)-(e) correspond to noise intensities higher than $D_{\mathrm{SR}}.$ The
dotted sinusoidal line drawn in the time domain represents the modulation
signal. }%
\label{timefrequencydomains}%
\end{center}
\end{figure}

Typical results of stochastic resonance measured in the time and the frequency
domains are shown at the left and right sides of Fig.
\ref{timefrequencydomains} respectively. Panels (a) and (b) correspond to low
noise levels below the critical value. Panel (a) shows the reflected
sinusoidal at $f_{\Omega}=1%
\operatorname{kHz}%
$ without jumps. Panel (b) shows the reflected sinusoidal containing a few
arbitrary jumps. Whereas, panel (c) which corresponds to the critical
stochastic noise $D_{\mathrm{SR}}$ exhibits one jump in the reflected signal
at every half cycle. That is, the \textit{time-scale matching condition} for
stochastic resonance given by $\tau(D_{\mathrm{SR}})=T_{\Omega}/2,$ is
fulfilled for this noise intensity, where $T_{\Omega}=2\pi/\Omega$, and
$\tau(D)$ is the metastable state lifetime corresponding to the noise
intensity $D.$ A possible theoretical linkage between the transition rate
$1/\tau$ exhibited by these resonators and their physical nonlinear mechanism
will be established in a future publication \cite{escape rate}. In panels (d)
and (e) on the other hand, the case of noise levels higher than
$D_{\mathrm{SR}}$ are shown. In panel (d), the coherence between the
modulating drive and the noise is lost, and multiple jumps are observed during
each cycle. Finally, in panel (e), the high noise almost screens the signal
and the induced jumps are totally uncorrelated. It is also apparent that the
reflected modulated signals shown in panels (c) and (d), in addition to being
amplified, exhibit a rather rectangular shape. Such distortion to the
sinusoidal shape at the output may originate from a dispersive character of
the system response as was suggested in Refs. \cite{Amp dist,Dykman}.

In the frequency domain displayed at the right side of panels (a)-(e), the
fundamental and the harmonics up to $n=5$ are shown, as the noise intensity is
increased. The appearance of even harmonics in the frequency spectrum
indicates that the nonlinear system is asymmetric, in contrast to symmetric
potential systems where only the odd harmonics emerge
\cite{Amplification,Higher-order}.

Furthermore, if one defines a spectral amplification of the \textit{n}-th
harmonic spectral density relative to that of the fundamental measured for the
noiseless case, one gets%

\begin{equation}
\eta_{n}(D)=A_{n}^{r}\left(  D\right)  /A_{1}^{r}\left(  D=0\right)  ,
\label{eta}%
\end{equation}
where the amplitude $A_{1}^{r}\left(  D=0\right)  $ is assumed to be
proportional to the input power of the modulation signal up to some constant
reflection factor. Since the different harmonics of the signal are strongly
correlated, we choose in the following figures to exhibit the amplification of
the third harmonic instead of the fundamental one $n=1.$ This is mainly
because the first harmonic resides at the frequency of the modulation source
signal and part of it bypasses the resonator.%
\begin{figure}
[pb]
\begin{center}
\includegraphics[
height=2.6351in,
width=3.4566in
]%
{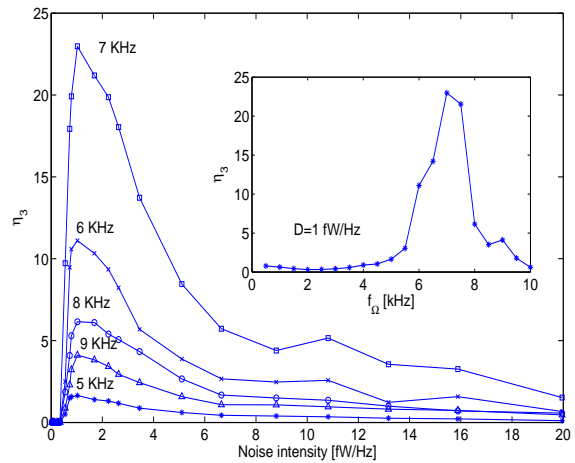}%
\caption{Third harmonic amplification $\eta_{3}$ as defined in Eq. \ref{eta}
plotted as a function of input noise intensity. The different plots correspond
to $f_{\Omega}=5$, $6$, $7$, $8$, $9$ $\operatorname{kHz}$ modulation
frequencies. The inset shows $\eta_{3}$ gain as a function of $f_{\Omega},$
corresponding to a constant input noise intensity of $1$ f$\operatorname{W}%
/\operatorname{Hz}.$}%
\label{sta3frequencyparameter}%
\end{center}
\end{figure}

In Fig. \ref{sta3frequencyparameter} the amplification parameter $\eta_{3}$ of
the third harmonic as a function of the input noise intensity was measured at
a fixed modulation amplitude but for different AM drive frequencies
$f_{\Omega}=5$, $6$, $7$, $8$, $9$ $%
\operatorname{kHz}%
$. As it is apparent from the figure $\eta_{3},$ amplification is
non-monotonic upon increasing the forcing frequency $f_{\Omega}.$ The
amplification graphs rise up with frequency up to $7%
\operatorname{kHz}%
$ and afterwards decay down. A cross section of $\eta_{3}$ parameter at a
constant noise intensity of $1$ f$%
\operatorname{W}%
/%
\operatorname{Hz}%
$ is shown in the inset as a function of $f_{\Omega}$ varying in the range
$0.5-10%
\operatorname{kHz}%
.$ A peak in $\eta_{3}$ is observed at $f_{\Omega}=7.5$ $%
\operatorname{kHz}%
.$ Similar amplification behavior was observed for the fundamental and for the
fifth harmonic as well. However, this non-monotonic behavior of $\eta_{3}$ as
a function of the forcing frequency somehow departs from the monotonic
decrease presented for example in Refs. \cite{SR review,Amplification}. This
may be due to the nature of the nonlinearity governing our system
\cite{dynamics} which can not be explained in terms of Duffing oscillator
nonlinearity, that was assumed in the cited papers. A theoretical prediction
of a non-monotonic SNR versus $f_{\Omega}$ was introduced though in Ref.
\cite{piecewise}, where the case of two square potential wells separated by a
square barrier was considered.%

\begin{figure}
[ptb]
\begin{center}
\includegraphics[
height=2.5763in,
width=3.8216in
]%
{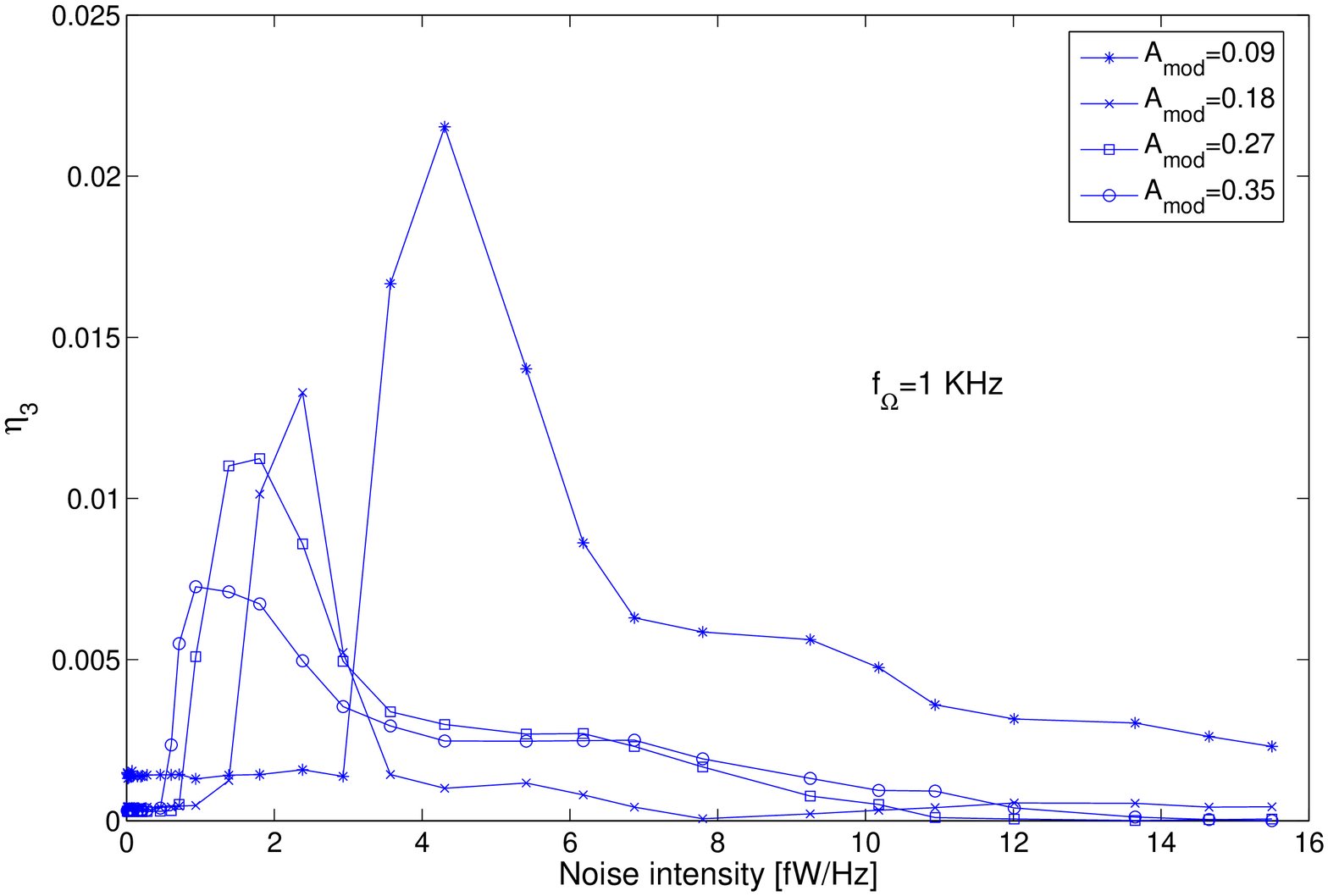}%
\caption{Third harmonic amplification $\eta_{3}$ as defined in Eq. \ref{eta}
plotted as a function of input noise intensity at $f_{\Omega}%
=1\operatorname{kHz}$. The different plots correspond to an increasing
modulation amplitude $A_{\operatorname{mod}}=0.09,$ $0.18,$ $0.27,$ $0.35$. }%
\label{eta3sensitivityparameter}%
\end{center}
\end{figure}

In contrast, when the modulation amplitude is increased at a fixed forcing
frequency, the resultant amplification is expected to decrease, mainly due to
increased nonlinear effects at higher modulation amplitudes, which tend to
suppress stochastic resonance phenomenon \cite{Amplification}. A good
confirmation to this hypothesis is presented in Fig.
\ref{eta3sensitivityparameter}, which displays decaying amplification curves,
as a function of the noise intensity, upon increasing the modulation
amplitudes $A_{\operatorname{mod}}=0.09,$ $0.18,$ $0.27,$ $0.35$ at a fixed
forcing frequency of $f_{\Omega}=1%
\operatorname{kHz}%
.$

In conclusion, nonlinear NbN superconducting resonators have been shown to
exhibit stochastic resonance when driven into the bistable region.
Amplification of a slowly varying AM signal carried by a microwave pump can be
achieved by establishing a resonant cooperation between the modulating signal
and the injected stochastic noise. Such amplification scheme may be useful in
the detection and amplification of small AM signals in communication area.
Moreover, stochastic resonance phenomenon can be employed to some extent, to
"probe" some important characteristics of the nonlinearity mechanism of the
resonators \cite{escape rate}.

This work was supported by the German Israel Foundation under grant
1-2038.1114.07, the Israel Science Foundation under grant 1380021, the Deborah
Foundation, the Poznanski Foundation, and MAFAT.

\bibliographystyle{plain}
\bibliography{attachfile}

\begin{thebibliography}{99}                                                                                               %


\bibitem {SR review}L. Gammaitoni, P. H\"{a}nggi, P. Jung, and F. Marchesoni,
Rev. Mod. Phys. \textbf{70}, 223 (1998).

\bibitem {Tuning in to noise}A. R. Bulsara and L. Gammaitoni, Physics Today p.
39 (March 1996).

\bibitem {Bona Fide}L. Gammaitoni, F. Marchesoni, and S. Santucci, Phys. Rev.
Lett. \textbf{74}, 1052 (1995).

\bibitem {Amplification}P. Jung, and P. H\"{a}nggi, Phys. Rev. A \textbf{44},
8032 (1991).

\bibitem {Amp dist}M. Morillo and J. G\'{o}mez-Ord\'{o}\~{n}ez, Phys. Rev. E
\textbf{51}, 999 (1995).

\bibitem {Benzi SR}R. Benzi, A. Satera, G. Parisi, and A. Vulpiani, J. Phys. A
\textbf{14}, L453 (1981).

\bibitem {Benzi Ice}R. Benzi, A. Satera, G. Parisi, and A. Vulpiani, SIAM J.
Appl. Math. \textbf{43}, 565 (1983).

\bibitem {schmitt}S. Fauve and F. Heslot, Phys. Lett. A \textbf{97}, 5 (1983).

\bibitem {neuronal}A. Longtin, A. Bulsara, and F. Moss, Phys. Rev. Lett.
\textbf{67}, 656 (1991).

\bibitem {cricket}J. E. Levin and J. P. Miller, Nature \textbf{380}, 165 (1996).

\bibitem {ring}B. McNamara, K. Wiesenfeld, and R. Roy, Phys. Rev. Lett.
\textbf{60}, 2626 (1988).

\bibitem {flux}R. Rouse, S. Han, and J. E. Lukens, Appl. Phys. Lett.
\textbf{66}, 108 (1995).

\bibitem {JJloop}A. D. Hibbs, A. L. Singsaas, E. W. Jacobs, A. R. Bulsara, J.
J. Pekkedahl, and F. Moss, J. Appl. Phys. \textbf{77}, 2582 (1995).

\bibitem {Beams}R. L. Badzey and P. Mohanty, Nature \textbf{437}, 995 (2005).

\bibitem {observation}B. Abdo, E. Arbel-Segev, O. Shtempluck, and E. Buks,
cond-mat/0501114 to be published in IEEE Trans. Appl. Supercond.

\bibitem {dynamics}B. Abdo, E. Arbel-Segev, O. Shtempluck, and E. Buks, Phys.
Rev. B \textbf{73}, 134513 (2006).

\bibitem {escape rate}B. Abdo, E. Arbel-Segev, O. Shtempluck, and E. Buks,
manuscript in preparation.

\bibitem {Dykman}M. I. Dykman, D. G. Luchinsky, R. Mannella, P. V. E.
McClintock, N. D. Stein, and N. G. Stocks, J. Stat. Phys. \textbf{70}, 463 (1993).

\bibitem {Higher-order}M. E. Inchiosa, A. R. Bulsara, and L. Gammaitoni, Phys.
Rev. E \textbf{55}, 4049 (1997).

\bibitem {piecewise}V. Berdichevsky and M. Gitterman, J. Phys. A \textbf{29},
L447 (1996).
\end{thebibliography}

\end{document}